\documentclass[iop]{emulateapj}

\usepackage{color}
\usepackage{graphicx}
\usepackage{tabularx}
\usepackage{placeins}
\usepackage{amsmath}
\usepackage{natbib}
\usepackage[utf8]{inputenc}

\newcommand{\kms}{\,km\,s$^{-1}$}

\shorttitle{Updated System Parameters of WASP-55\textup{b} and WASP-75\textup{b}}
\shortauthors{B. J. M. Clark et al.}

\begin{document}
\title{An Analysis of Transiting Hot Jupiters Observed with K2: WASP-55\lowercase{b} and WASP-75\lowercase{b} }

\author{B. J. M. Clark\altaffilmark{1},
	D. R. Anderson\altaffilmark{1},
	C. Hellier\altaffilmark{1},
	O. D. Turner\altaffilmark{2} and
	T. Mo{\v c}nik\altaffilmark{1}
}
\affil{\textsuperscript{1}Astrophysics Group, Keele University, Staffordshire, ST5 5BG, UK}
\affil{\textsuperscript{2}Observatoire de Genève, Université de Genève, 51 Chemin des Maillettes, 1290 Sauverny, Switzerland}


\begin{abstract}
We present our analysis of the K2 short-cadence data of two previously known hot Jupiter exoplanets: WASP-55\lowercase{b} and WASP-75\lowercase{b}.
The high precision of the K2 lightcurves enabled us to search for transit timing and duration variations, rotational modulation, starspots, phase-curve variations and additional transiting planets. 
We identified stellar variability in the WASP-75 lightcurve which may be an indication of rotational modulation, with an estimated period of $11.2\pm1.5$ days.
We combined this with the spectroscopically measured $v\sin(i_*)$ to calculate
 a possible line of sight projected inclination angle
 of $i_*=41\pm16^{\circ}$.
We also perform a global analysis of K2 and previously published data to refine the system parameters.
\end{abstract}



\section{Introduction} \label{sec:intro}
The K2 mission has been in operation since May 2014 following the failure of two reaction wheels in the original Kepler mission \citep{2014PASP..126..398H}, monitoring fields along the ecliptic plane.
The spacecraft experiences a roll which introduces systematics in the photometric lightcurves on a time-scale of approximately 6 hr when it is corrected by a thruster event.
As a result the quality of the photometry can be degraded by up to a factor of 4 \citep{2014PASP..126..398H}.
Despite this, methods such as self flat-fielding (SFF)(\citealt{2014PASP..126..948V}), K2SC \citep{2016MNRAS.459.2408A} and routines that we have built \citep{2016AJ....151..150M} can reduce this effect to the extent that the quality of the produced data is near that of the original mission.

Following the initial discovery of an exoplanet, subsequent observations are common if a planet possesses unique characteristics (e.g. WASP-18b -- \citealt{2009Natur.460.1098H}).
They are also frequent if the system is readily observable from Earth; for example, those with a bright host star such as WASP-33b \citep{2006MNRAS.372.1117C} or planets with a large transit depth such as WASP-43b \citep{2011AA...535L...7H}.
For other planets however, follow-up observations can range from infrequent to absent, meaning we can miss crucial discoveries about these systems. An example of this are the recent K2 observations of the WASP-47 system that revealed two additional companions \citep{2015ApJ...812L..18B}. 
WASP-55b and particularly WASP-75b are examples of two planets that have had little follow-up since their initial discoveries.

WASP-55b was found to be a moderately inflated hot Jupiter with a mass of $0.57 \pm 0.04$ $M_{\textrm{Jup}}$ and a radius of $1.30 \pm 0.05$ $R_{\textrm{Jup}}$ orbiting a G1 star discovered in 2011 \citep{2012MNRAS.426..739H}.
\begin{figure}[t]
	\centering
	\includegraphics[width=0.45\textwidth]{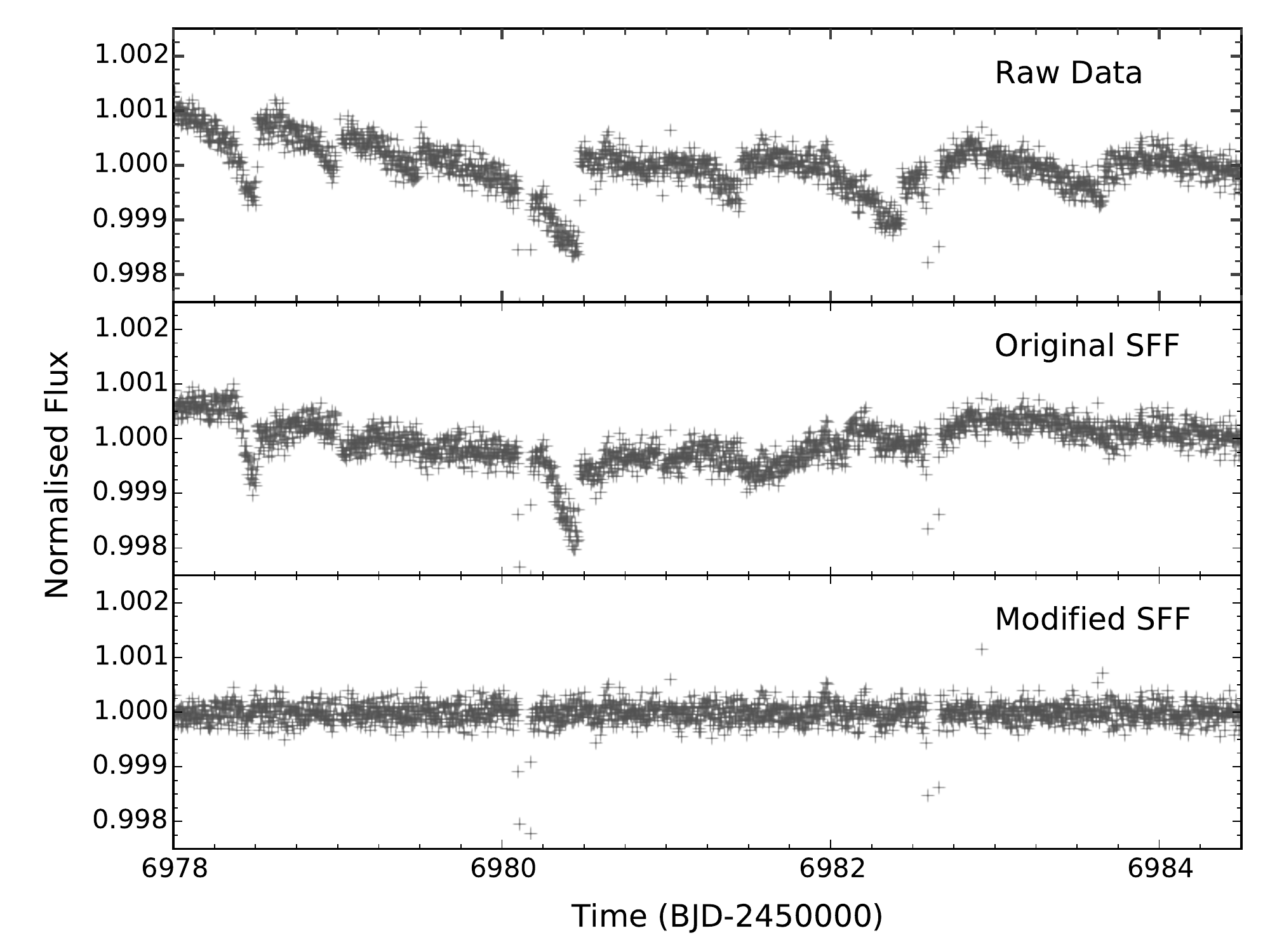}
	\caption{\label{fig:detrendcompare}  The top panel shows a section of the raw WASP-75b lightcurve that contains strong systematics that correlate with PSF position, binned to 5 minute intervals. 
		The middle panel shows the same section of the de-trended lightcurve produced by the original self-flat fielding method and the bottom panel shows the de-trended lightcurve produced by the modified method used in this paper.
	} 
\end{figure}
WASP-75b was discovered by \citet{2013AA...559A..36G} as a hot Jupiter with a mass of $1.07 \pm  0.05$ $M_{\textrm{Jup}}$ and a radius of $1.27\pm 0.05$ $R_{\textrm{Jup}}$. It orbits WASP 75, a F9 star with an orbital period of 2.484 days.
In this paper we present a refined set of system parameters for WASP-55b and WASP-75b.
We also search the K2 lightcurves for transit timing and duration variations, 
stellar rotational modulation, starspot occultations, phase-curve variations and search the 
residual lightcurves for additional transiting companions.

\begin{figure*}[!t]
	\centering
	\includegraphics[width=1\textwidth]{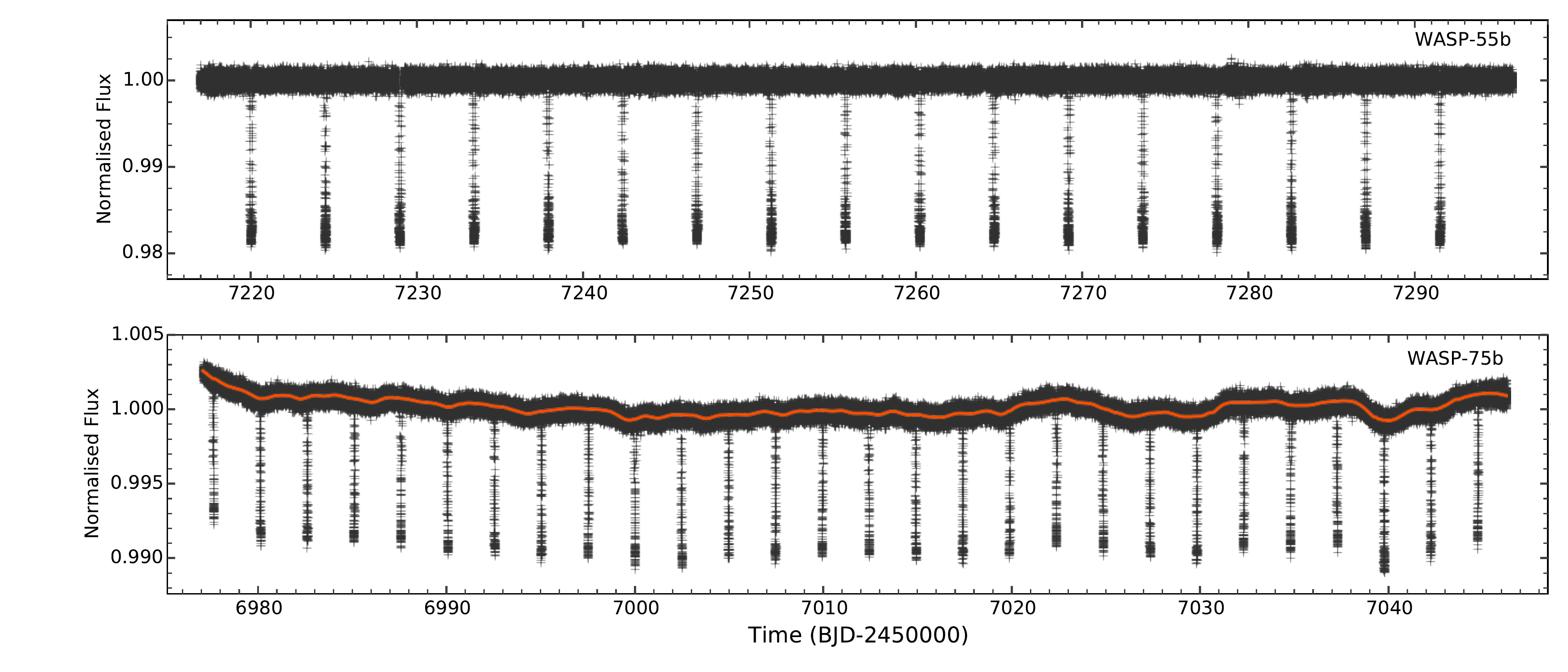}
	\caption{\label{fig:lightcurves} The top plot shows the final lightcurve of WASP-55b obtained by de-trending and clipping the raw lightcurve as described in Section \ref{sec:detrending}. The bottom plot shows the SFF de-trended and clipped lightcurve for WASP-75b. The orange line represents the detected low-frequency stellar variation that is removed before performing the global Markov chain Monte Carlo (MCMC) analysis in Section \ref{sec:params}. } 
	
\end{figure*}

\section{K2 Data Reduction} \label{sec:reduction}
\subsection{Data Extraction}\label{sec:extraction}
WASP-55b (EPIC 212300977) was observed during campaign 6 of the K2 mission, which ran from 2015 July 14 until 2015 September 30. It produced a total of 112,672 short-cadence images.
WASP-75b (EPIC 206154641)was observed during campaign 3, it produced 101,370 short-cadence images between 2014 November 14 and 2015 February 3.
We retrieved the target pixel files for each system using the Barbara A.
Mikulski Archive for Space Telescopes (MAST\footnote{https://archive.stsci.edu/k2/}).

\begin{figure}[t]
	\centering
	\includegraphics[width=0.5\textwidth]{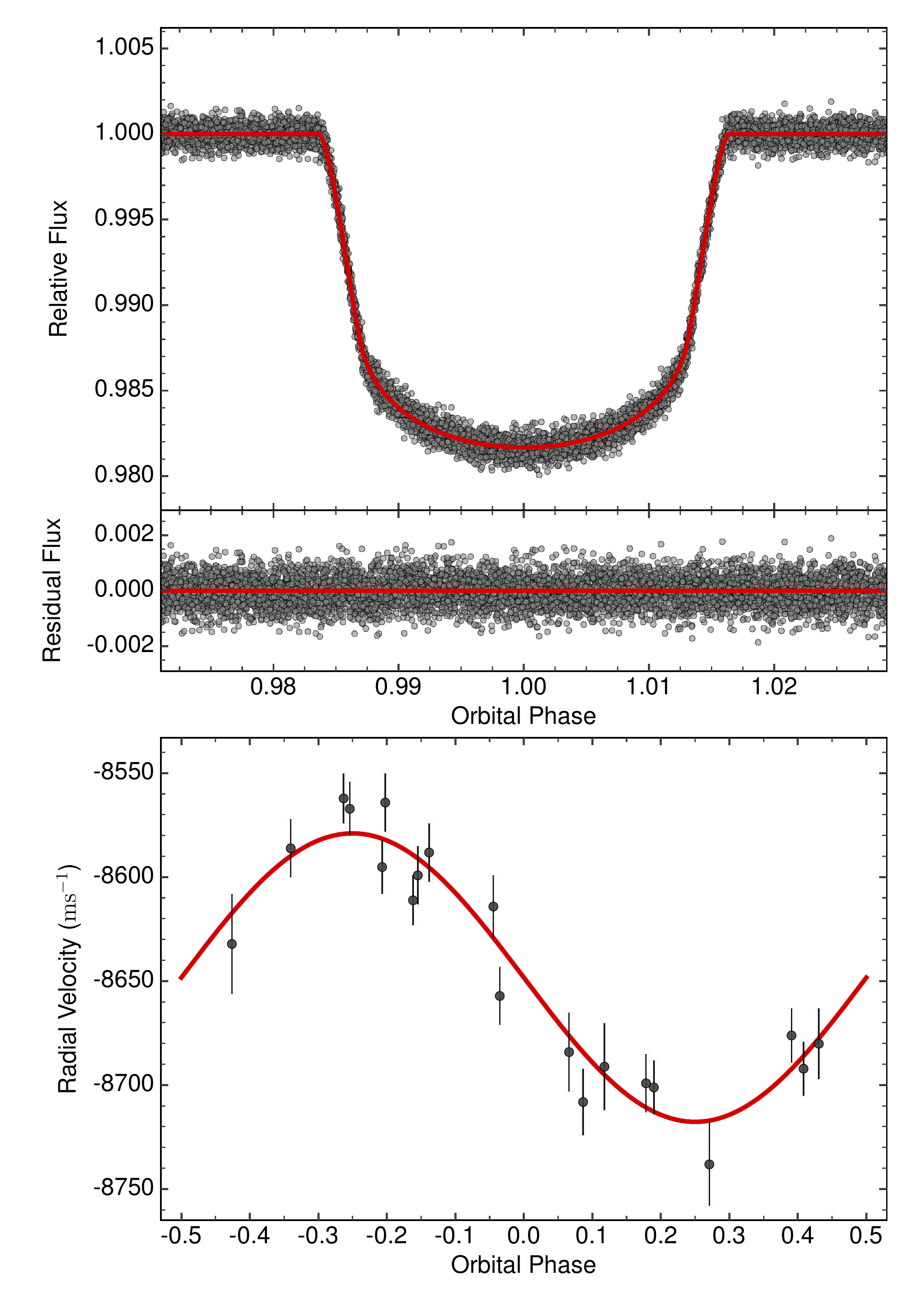}
	\caption{\label{fig:w55phaselc} The top plot shows the fully de-trended, phase-folded K2 lightcurve for WASP-55b. The model produced by the global MCMC run is shown by the red line.
		The middle panel shows the residuals of the fit and the bottom panel shows the radial velocity measurements with the best-fitting orbital model in red.}
\end{figure}

The large motion of the point spread function (PSF) on the detector in campaign 3 was cause for concern when reducing the raw images.
A traditional fixed-mask method used by a majority of the K2 data reduction to date (e.g. \citealt{2016AJ....151..150M}) appeared to degrade the precision of the output lightcurves.
To resolve this issue, we used an aperture photometry routine written using \textsc{PyRAF}, with aperture sizes ranging from 0.5 to 8.5 pixels in steps of 0.25 pixels that used a flux-weighted centroid method to re-position the aperture based upon the PSF position in each frame.
For the WASP-75b dataset an aperture size of 6.5 pixels led to a reduction in the RMS from 368 PPM to 325 PPM
and for WASP-55b, a 5.5 pixel aperture reduced the scatter from 568 PPM to 530 PPM.

\begin{figure}[t]
	\centering
	\includegraphics[width=0.5\textwidth]{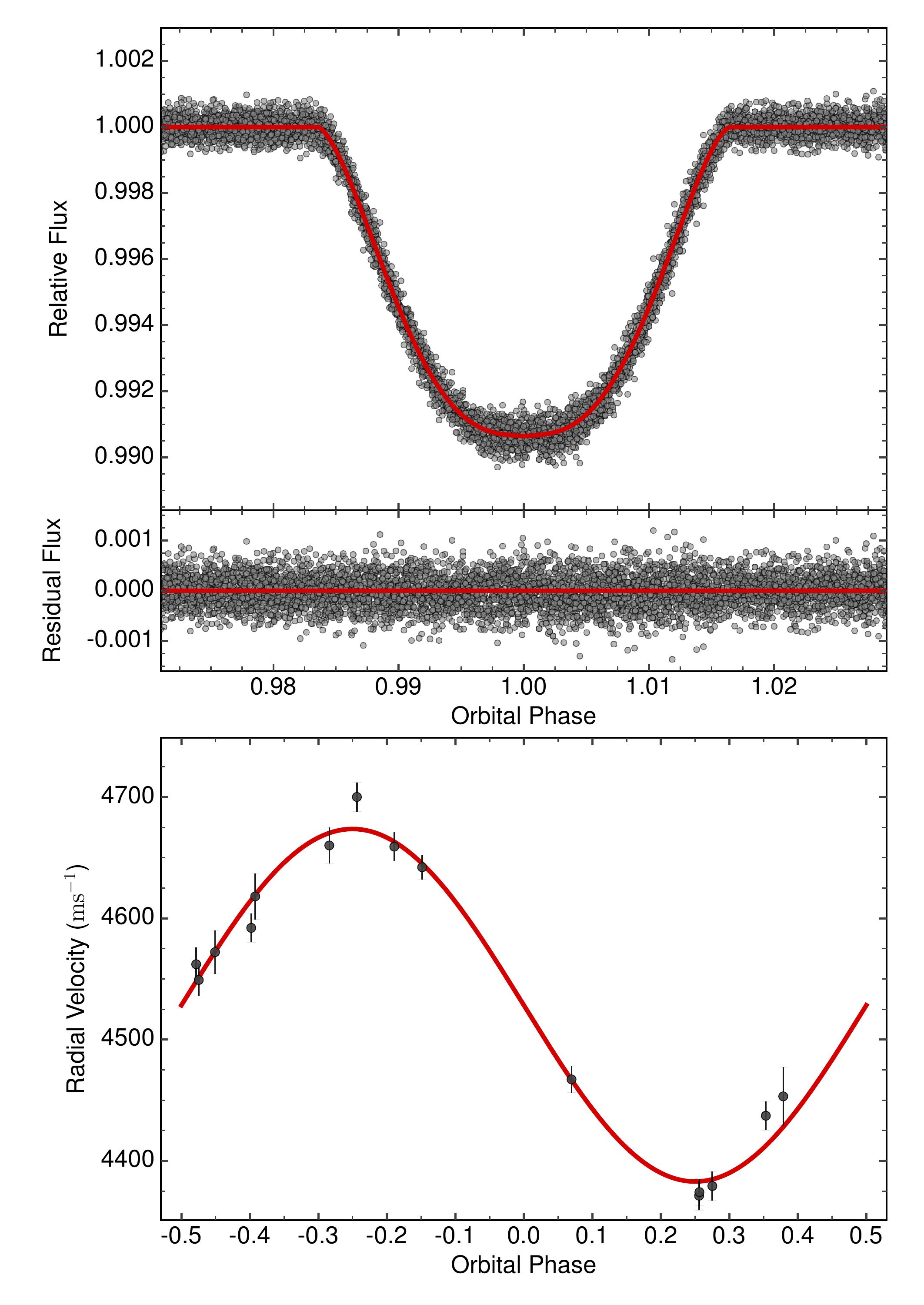}
	\caption{\label{fig:w75phaselc} The same as Figure \ref{fig:w55phaselc} except we plot WASP-75b. The top panel shows the fully de-trended, phase-folded K2 lightcurve and model in red. The middle panel shows the residuals of the fit and the bottom panel shows the radial velocity measurements with the best-fitting model in red.}
\end{figure}

\begin{table}[t]
\tiny
\caption{Orbital, stellar and planetary parameters from the MCMC analysis of WASP-55b (EPIC 212300977). 
We list each of the proposal parameters, derived parameters, and 
parameters controlled by priors separately. We tabulate our results and those provided by the literature. 
}
\label{tab:w55mcmcparams}
\begin{center}
\setlength\extrarowheight{4pt}
\begin{tabular}{lp{8em}p{8em}p{8em}}
Symbol (unit)                                  & This Work                                 & \citet{2012MNRAS.426..739H}          & \citet{2016MNRAS.457.4205S}       \\ \hline
\multicolumn{4}{l}{MCMC proposal parameters }                                                         \\ \hline
$P$ (days)                                        & $4.465630\pm0.000001$                     & $4.465633 \pm 0.000004$                & $4.4656291\pm0.0000011$             \\
$T_{\rm c}$ (days)                                & $7256.25436\pm0.00003$                    & $5737.9396 \pm 0.0003$                 & $6416.71565\pm0.00013$              \\
$T_{\rm 14}$ (days)                               & $0.1459\pm0.0002 $     & $0.147 \pm 0.001$                      & $0.147\pm0.003^a$             \\
$R_{\rm P}^{2}$/R$_{*}^{2}$                    & $0.01551\pm0.00005$  & $0.0158 \pm 0.0003$                    & $0.0155\pm0.0002$                   \\
$b$                                            & $0.18\pm0.03$           & $0.15 \pm 0.12$                        & $0.03\pm0.23^a$               \\
$K_{\rm 1}$ (m s$^{-1}$)                       & $69\pm4$                                  & $70 \pm 4$                             & $ 70 \pm 4$                         \\
$\gamma$ (m s$^{-1}$) \medskip                 & $-4324\pm3$                               & $-4324.4 \pm 0.9$                      & -                                   \\ \hline
\multicolumn{4}{l}{MCMC proposal parameters constrained by priors }                                   \\ \hline
$M_{\rm *}$ ($M_{\rm \odot}$)                  & $1.16\pm0.03$                             & $ 1.01 \pm 0.04$                       & $1.162\substack{ +0.029\\ -0.033}$  \\
$T_{\rm eff}$ (K)                              & $6070\substack{+51 \\ -46}$               & $5960 \pm 100$                         & $6070 \pm 53$                       \\
Fe/H                             & $0.09 \pm 0.05$                           & $-0.20 \pm 0.08$                       & $0.09 \pm 0.05$                     \\ \hline
\multicolumn{4}{l}{MCMC derived parameters }                                     \\ \hline
$i$ ($^\circ$) \medskip                        & $89.0\pm0.2$             & $89.2 \pm 0.6$                         & $89.83\substack{+0.57\\ -1.20}$     \\
$e$                                            & $0$ (fixed)  (\textless0.22 at $3\sigma$) & 0 (fixed) (\textless0.20 at $3\sigma$) & -                                   \\
$a$ (au)                                       & $0.0558\pm0.0006$     & $0.0533 \pm 0.0007$                    & $ 0.0558\pm0.0005$                  \\
$R_{\rm *}$ ($R_{\rm \odot}$)                  & $1.11\pm0.01$                             & $1.06\substack{+0.03\\-0.02}$          & $ 1.102\substack{+0.020\\-0.015}$   \\
$\log g_{*}$ (cgs)                             & $4.413\pm0.006$                           & $4.39\substack{+0.01\\ -0.02}$         & $ 4.419\substack{ +0.009\\-0.015}$  \\
$\rho_{\rm *}$ ($\rho_{\rm \odot}$)            & $0.85\pm0.01$                             & $0.85\substack{+0.03 \\-0.07}$         & $0.869\substack{+0.026\\ -0.041}$   \\
$M_{\rm P}$ ($M_{\rm Jup}$)                    & $0.62\pm0.04$                             & $0.57 \pm 0.04$                        & $ 0.627\substack{ +0.037 \\-0.038}$ \\
$R_{\rm P}$ ($R_{\rm Jup}$)                    & $1.34\pm0.01$                             & $ 1.30\substack{+0.05 \\-0.03}$        & $ 1.335\substack{ +0.031 \\-0.020}$ \\
$\log g_{\rm P}$ (cgs)                         & $2.9\pm0.03$                              & $2.89 \pm 0.04$                        & $ 2.94\pm0.03$                      \\
$\rho_{\rm P}$ ($\rho_{\rm J}$)                & $0.26\pm0.02$                             & $ 0.26\substack{+0.02 \\-0.03}$        & $0.247\substack{ +0.017 \\-0.021}$  \\
$T_{P}$ (K)                                    & $1305\pm12^b$               & $1290 \pm 25$                          & $ 1300\substack{+15\\-13}$          \\ \hline
\end{tabular}

\tiny

\end{center}
{\scriptsize $^a$ Calculated using the parameters from \citet{2016MNRAS.457.4205S} using the equations of \citet{2003ApJ...585.1038S}\\
$^b$Assuming a zero bond albedo and efficient day--night redistribution of heat.\\}
\normalsize
\end{table}

\subsection{De-trending}\label{sec:detrending}
As noted in Section \ref{sec:intro}, K2 lightcurves contain systematics that correlate with position of stellar flux on the detector.
This is visibly seen as a sawtooth-like pattern in the lightcurve, which can be observed in Figure \ref{fig:detrendcompare}.
We were able to correct for a majority of these systematics using the methods of \citet{2016AJ....151..150M}.
However, for WASP-75b there were areas of strong systematics caused by a high spacecraft jitter, that were not corrected well by this method.
The systematics correlated heavily with the X and Y position of the PSF on the detector, with large jumps in position and flux at every thruster event.
We used a moving gradient to detect the areas with jumps and used these dates as boundaries between windows over which to correct the systematics, rather than a fixed window size.
With this method, a third order polynomial fit to the flux vs. arclength trend was enough to successfully remove the visible trends (Figure \ref{fig:detrendcompare}).

To model the low-frequency variability from the lightcurve of WASP-75b we used a Gaussian convolution method, similar to that of \citep{2016AJ....151..150M} but with a kernel size larger than the time-scales of systematic noise and transit events. This was then removed from the lightcurve before further analysis.
We also performed a running median filter, with a kernel size of 21 points, to clip all data that were greater than 8-$\sigma$ from the median-filtered residuals.
In total, we clipped 2409 and 4905 points from the WASP-55b and WASP-75b lightcurves respectively.
The de-trended lightcurves are shown in figure \ref{fig:lightcurves}.

\section{System Parameters}\label{sec:params}
To determine the parameters of the system, we used an adaptive Markov chain Monte Carlo (MCMC) routine \citep{2007MNRAS.380.1230C,2008MNRAS.385.1576P,2015AA...575A..61A}
to simultaneously analyse the K2 lightcurves with their respective, previously published radial velocity (RV) data.

For WASP-55b and WASP-75b, we used the normalised K2 lightcurves with the \small{CORALIE} RVs from  \citet{2012MNRAS.426..739H} and \citet{2013AA...559A..36G}, respectively.
We assumed that the orbit was circular for the main MCMC runs, but set eccentricity as a free parameter on subsequent runs to place a constraint on its upper limit for both systems.
We performed an additional run using the ground-based transit lightcurves from \citet{2012MNRAS.426..739H} for WASP-55b and \citet{2013AA...559A..36G} for WASP-75b to refine the ephemeris of both planets by extending the baseline.
We present the updated system parameters in Tables \ref{tab:w55mcmcparams} and \ref{tab:w75mcmcparams} and the phase-folded lightcurves and models in Figures \ref{fig:w55phaselc} and \ref{fig:w75phaselc}.

We used a  four-parameter law to determine the limb darkening coefficients with values interpolated from those of \citet{SING2010} and based upon stellar temperature.
We found that, for both systems, the stellar effective temperatures produced limb darkening coefficients that were in good agreement with the shape of the lightcurve.
There were no visible anomalies in the residuals (see the middle panel in Figures \ref{fig:w55phaselc} and \ref{fig:w75phaselc})
as has been the case for some planets that we have studied (e.g. \citealt{2017MNRAS.471..394M}).

We obtained our values of the stellar mass from a comparison with stellar models by using the \textsc{BAGEMASS} code of \citet{2015AA...575A..36M}.
This took as inputs, the spectroscopic values of stellar effective temperature and metallicity 
($\lbrack \frac{\textrm{Fe}}{\textrm{H}} \rbrack$) as well as stellar density from initial MCMC runs.
We used the calculated values as a prior constraint in our global MCMC.

\citet{2016AA...589A..58E} discovered a faint, nearby companion to WASP-55b.
They determined that it had a magnitude difference of $5.210\pm0.018$ in the $r_{\textrm{\tiny{TCI}}}$ band at a distance of $4\farcs345\pm0\farcs010$, placing it within our aperture.
We used Equation 3 of \citet{2009A&A...498..567D} to correct our lightcurve for the additional flux measured from the companion star. 
The result was a minor difference in the calculated eclipse depth, we find a new value of $0.01551\pm0.00005$ compared to $0.01550\pm0.00004$ before the correction.


\section{No Transit duration or timing variations}

Additional planetary companions can cause variations in the timing of transit events due to the gravitational perturbations that they cause (\citealt{2005MNRAS.359..567A,2005Sci...307.1288H}).
We searched for transit timing variations (TTVs) and Transit duration variations (TDVs) in the WASP-55b and WASP-75b datasets by splitting the lightcurves at midpoints between each transit. 
We then performed a single MCMC run for each transit, with no other input data. 
We used the parameters from the global MCMC run as prior constraints, with transit epoch and duration set as free parameters.

\begin{table}[t]
\tiny
\caption{In the same way as Table \ref{tab:w55mcmcparams} we show the orbital, stellar and planetary parameters from the MCMC analysis of WASP-75b (EPIC 206154641), listing each of the proposal, derived and 
parameters controlled by priors, separately. We again tabulate our results and those provided by the literature. 
}
\label{tab:w75mcmcparams}
\begin{center}
\setlength\extrarowheight{4pt}
\begin{tabular}{llp{11em}}
Symbol (unit)                                  & This Work                                & \citet{2013AA...559A..36G}        \\ \hline
\multicolumn{3}{l}{MCMC proposal parameters}                                \\ \hline
$P$ (days)                                        & $2.4842014\pm0.0000004$                  & $2.484193 \pm 0.000003$             \\
$T_{\rm c}$ (days)                                & $7009.94594\pm0.00002$                   & $ 6016.2669 \pm 0.0003$             \\
$T_{\rm 14}$ (days)                               & $0.08097\pm0.00008$ & $0.0822 \pm 0.0011$                 \\
$R_{\rm P}^{2}$/R$_{*}^{2}$                    & $0.01133\pm0.00005$ & $ 0.0107 \pm 0.0003$                \\
$b$                                            & $0.8926\pm0.0007$    & $0.882\substack{+0.006\\-0.008}$    \\
$K_{\rm 1}$ (m s$^{-1}$)                       & $145\pm4$                                & $146 \pm 4$                         \\
$\gamma$ (m s$^{-1}$) \medskip                 & $2264\pm3$                               & $ 2264.29 \pm 0.06$                 \\ \hline
\multicolumn{3}{l}{MCMC proposal parameters constrained by priors}                         \\ \hline
$M_{\rm *}$ ($M_{\rm \odot}$)                  & $1.16\pm0.03$                            & $1.14 \pm 0.07$                     \\
$T_{\rm eff}$ (K)                              & $6035\substack{+88 \\ -93}$              & $6100 \pm 100$                      \\
Fe/H & $0.07\pm0.09$                            & $0.07 \pm 0.09$                     \\ \hline
\multicolumn{3}{l}{MCMC derived parameters}                             \\ \hline
$i$ ($^\circ$) \medskip                        & $81.96\pm0.02$                           & $82.0\substack{+0.3\\-0.2}$         \\
$e$                                            & 0 (fixed) (\textless0.10 at $3\sigma$)   & $0$ (fixed)                         \\
$a$ (au)                                       & $0.0377\pm0.0006$                        & $0.0375\substack{+0.0007\\-0.0008}$ \\
$R_{\rm *}$ ($R_{\rm \odot}$)                  & $1.27\pm0.02$                            & $ 1.26 \pm 0.04$                    \\
$\log g_{*}$ (cgs)                             & $4.294\pm0.008$       & $ 4.29 \pm 0.02$                    \\
$\rho_{\rm *}$ ($\rho_{\rm \odot}$)            & $0.566\pm0.003$                          & $0.56 \pm 0.04$                     \\
$M_{\rm P}$ ($M_{\rm Jup}$)                    & $1.08\pm0.05$                            & $ 1.07 \pm 0.05$                    \\
$R_{\rm P}$ ($R_{\rm Jup}$)                    & $1.31\pm0.02$                            & $ 1.270 \pm 0.048$                  \\
$\log g_{\rm P}$ (cgs)                         & $3.16\pm0.01$                            & $3.179\substack{+0.033\\-0.028}$    \\
$\rho_{\rm P}$ ($\rho_{\rm J}$)                & $0.48\pm0.02$                            & $0.52\substack{+0.06\\-0.05}$       \\
$T_{P}$ (K)                                    & $1688\substack{+25 \\ -26}^a$              & $1710 \pm 20$                       \\ \hline
\end{tabular}

\tiny

\end{center}
$^a$Assuming a zero bond albedo and efficient day--night redistribution of heat.\\
\normalsize
\end{table}

Against the null hypothesis of equally spaced and equal duration transit events, for WASP-55b we found
a $\chi^2$ of 28.7 and 6.3 for the TTVs and TDVs, respectively, with 17 degrees of freedom.
We place upper limits on the TTVs of 25s and 100s for the TDVs.
For WASP-75b, the measured TTVs and TDVs had a $\chi^2$ of 24.8 and 36.3, with 28 degrees of freedom.
The upper limits for the TTVs and TDVs were 35 s and 120 s respectively.
Given the lack of significant TTVs and TDVs, we can rule out the existence of large, close-in companion planets for both systems. 

\begin{figure}[t]
	\centering
	\includegraphics[width=0.45\textwidth]{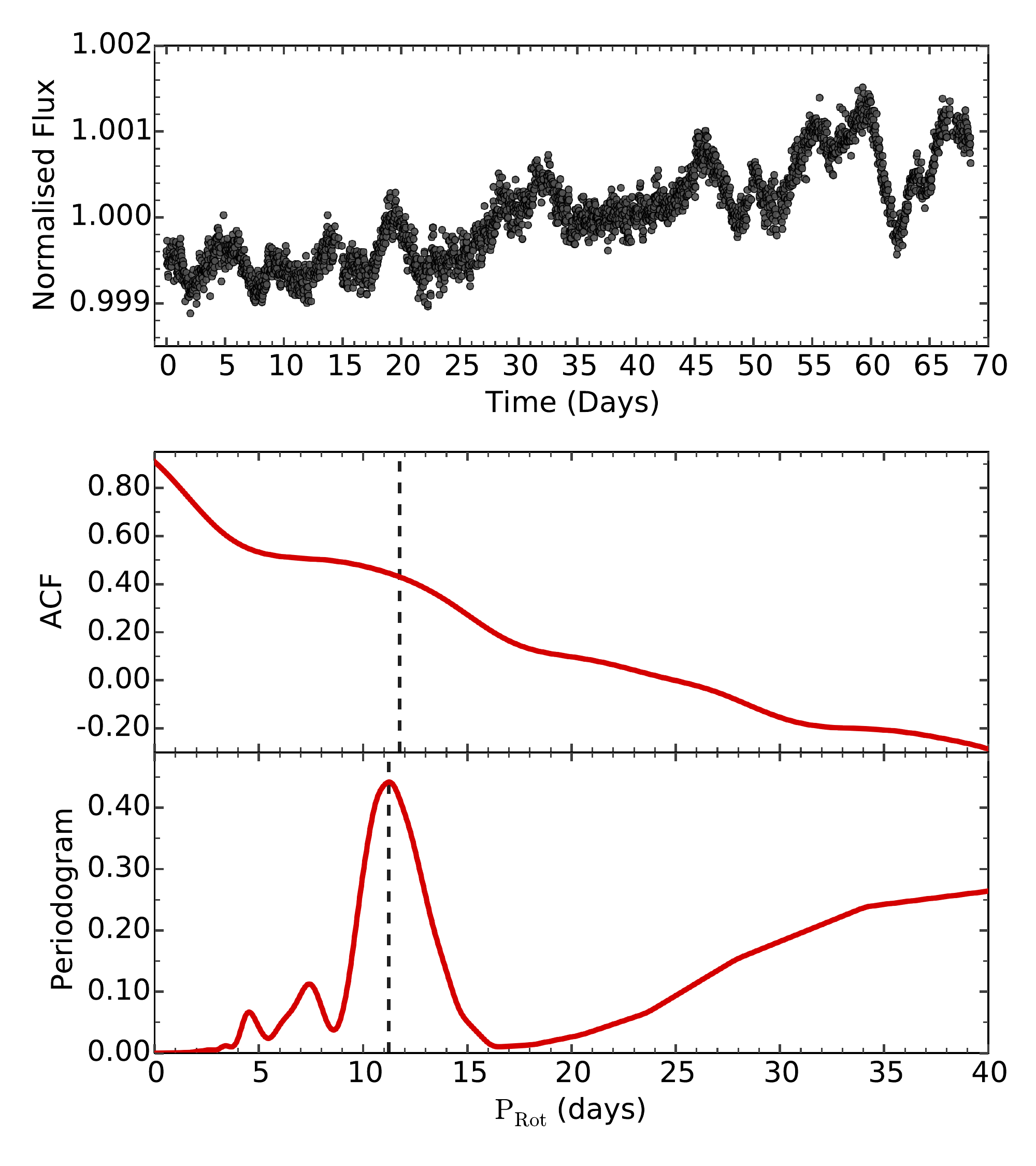}
	\caption{\label{fig:w75stellaramp}  The top plot shows the transit-subtracted, SFF de-trended, sigma-clipped WASP-75b lightcurve, binned to 10 minute intervals.
		The middle plot shows the ACF profile in red, with the center of the detected peak represented by a black dashed line. 
		The bottom plot displays the LS periodogram of the system, again with the strongest peak shown by the black vertical line.
	}
\end{figure}

\section{No starspot occultations but stellar variation in WASP-75}\label{sec:rotation}

We performed a thorough visual inspection for starspots in both the de-trended lightcurves and model-subtracted residuals but we found no evidence of starspot occultations in either the WASP-55b or WASP-75b data.

We find no variations in the WASP-55b lightcurve but do detect low-frequency variations in the WASP-75b lightcurve.
To investigate these, 
we used { \sc K2SC} \citep{2016MNRAS.459.2408A} to obtain a systematic-corrected lightcurve from the pre-search conditioned K2 data, which included low-frequency variations but excluded transit events (Figure \ref{fig:w75stellaramp}).
We used this method as our low-frequency variation lightcurve included a long-term trend that was removed well by the pre-search data conditioning module.
We used the autocorrelation function (ACF) of \citet{2013MNRAS.432.1203M} and a Lomb--Scargle (LS) periodogram to search the lightcurve for rotational modulation.
From the ACF, we found a period of $11.7\pm0.5$ days determined from the first three peaks, that is possibly indicative of rotational modulation.
This is in agreement with the value of $11.2\pm1.5$ produced by the LS-periodogram.

For WASP-75b, we calculated an updated value for macroturbulence ($v_{\textrm{mac}}$) of $4.05\pm0.41$ \kms using the calibrations of \citet{2014MNRAS.444.3592D} and produced a new value of $v\sin(i_*)=3.8\pm1.0$ \kms
Assuming spin-orbit alignment, this implies a stellar rotation period of $16.9\pm4.5$ days.
The marginal agreement between the predicted and measured values of stellar rotation could hint at the possibility of a non-aligned stellar inclination angle.
We used the new $v\sin(i_*)$ with the more conservative LS-periodogram measurement of the rotation period of the star to determine that WASP-75b has a possible rotation speed of $v=5.7\pm0.8$  \kms and stellar line-of-sight inclination angle of $i_*=41\pm16^{\circ}$.
If we assumed sun-like starspot latitudes and differential rotation, the stellar line-of-sight inclination angle would be $i_*=39\pm14^{\circ}$.

It is possible to use the value of $i_*$ with the obliquity angle, that can be measured using the Rossiter--McLaughlin (RM) effect
(\citealt{1924ApJ....60...15R,1924ApJ....60...22M,2017arXiv170906376T}), to calculate the true angle ($\Psi$) between the stellar rotation and orbital axes.
$\Psi$ is important in theories of the formation and evolution of planetary systems \citep{2016ApJ...819...85C}.
A RM measurement of WASP-75b would therefore be beneficial in this case. 
We estimated the amplitude of the RM effect to be $13\pm3$ m\,s$^{-1}$ for WASP-75b. 
This effect should be measurable with high-resolution spectrographs and we predict a typical RV precision of $\sim4$ m\,s$^{-1}$ from a 900 s HARPS spectrum of WASP-75.

\section{No phase-curve variations}\label{sec:phasevar}

At optical wavelengths, phase-curve variations are expected to comprise of four main constituents:
ellipsoidal variations \citep{2010ApJ...713L.145W,2012ApJ...751..112J}, doppler beaming \citep{2012ApJ...745...55G}, a component of reflected light from the star \citep{2012ApJ...747...25M} and a secondary eclipse \citep{2013ApJ...772...51E}.
Using equations from \citet{2010A&A...521L..59M} and the system parameters in Tables \ref{tab:w55mcmcparams} and \ref{tab:w75mcmcparams} we calculated the predicted amplitudes of these effects and created a phase-curve for each of the two systems.
\begin{figure}[t]
	\centering
	\includegraphics[width=0.45\textwidth]{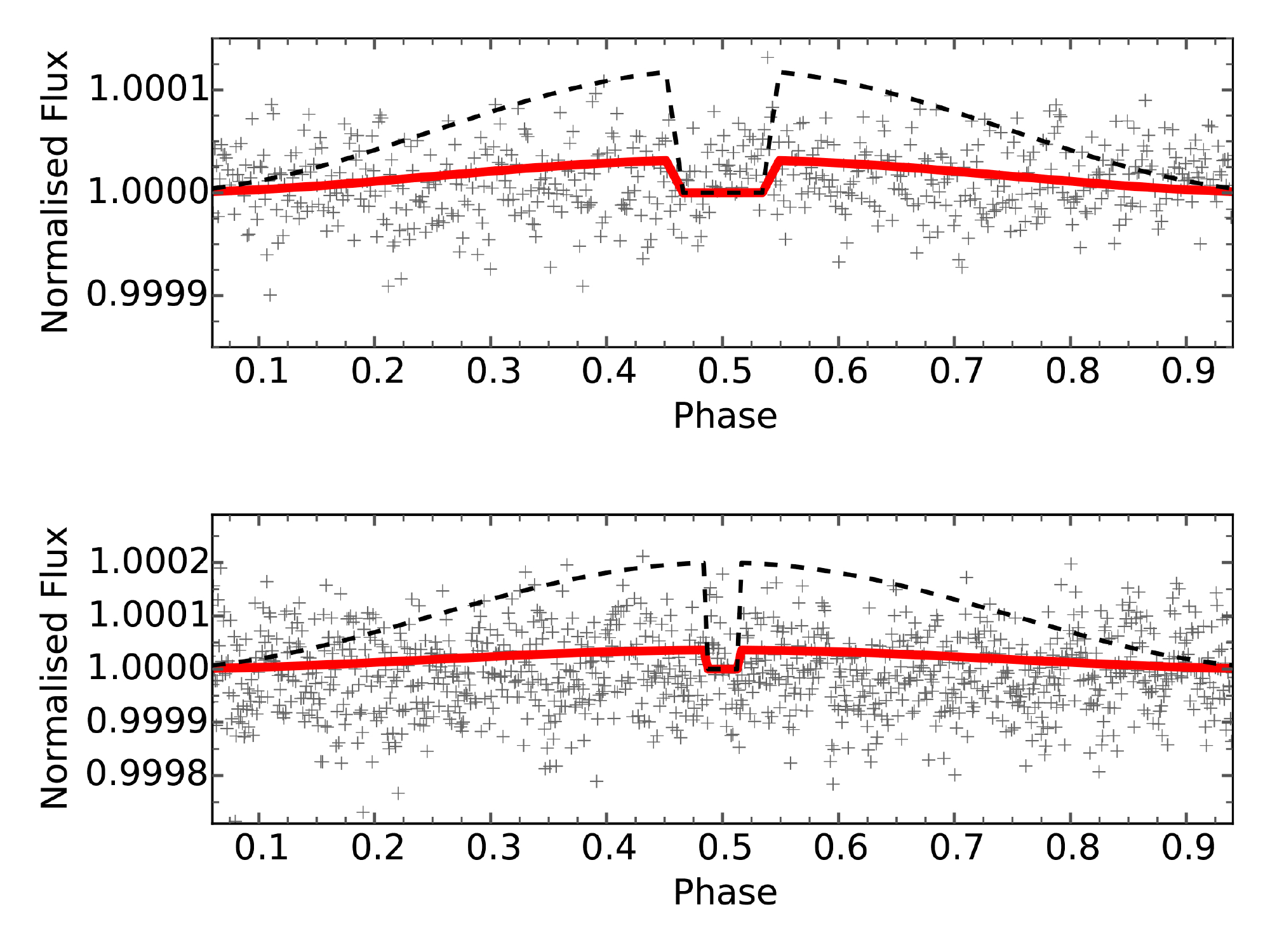}
	\caption{\label{fig:w55phase}  The phase-folded residuals for the WASP-75b (top) and WASP-55b (bottom) lightcurves, binned to 5 minute intervals.
		The red lines show the models for the predicted phase variations described in Section \ref{sec:phasevar} and the predicted secondary eclipses at an orbital phase of 0.5.
		The black dashed lines show the model for an estimated upper limit for phase variations in each system.
	}
\end{figure}
We calculated predicted values of the ellipsoidal variations, Doppler beaming, reflection and secondary eclipse depth as 0.4, 0.9 and 13 PPM for WASP-55b and 3.4, 1.9 and 26 PPM for WASP-75b.
 
We used a Levenberg--Marquardt algorithm to attempt to fit the above model to the phase-folded residual lightcurve of each system.
However, in both cases, we found a negligible amplitude provided the best fit, which was also confirmed visually.
We therefore did not detect any phase variations in either system.
The predicted values are lower than the precision of the lightcurves;
even if variation of this magnitude were to exist, we would not be able to detect them with the current data.
It is possible that the de-trending methods used could have removed phase variations from the original data, but signal injection tests performed by \citet{2017MNRAS.471..394M} have shown that the SFF method should preserve periodic variations.
We therefore placed conservative estimates on the upper limits of the phase curve variations and secondary eclipse depths of 100 PPM for WASP-55b and 60 PPM for WASP-75b.
As a result, we have placed upper limits on the geometric albedos of 0.8 and 0.2 for WASP-55b and WASP-75b, respectively. 
We plot the phase-folded, binned residual lightcurves in Figure \ref{fig:w55phase} along with the predicted and lower-limit phase-curve models.

\section{No Additional Transiting Planets}
To search for signals of additional transiting planets we used the box-least-squares method of \citet{2006MNRAS.373..799C} on the model-subtracted residuals for each planet.
We found no significant signals in the residuals of WASP-55b dataset but
did find significant peaks in the periodogram of WASP-75. 
However, further inspection revealed that this was due to the presence of residual correlated noise. 
We therefore found no signals from additional transiting planets with periods between 0.5 and 35 days in either the WASP-55b or WASP-75b data and we placed upper limits on the eclipse depths of additional planets to 280 and 190 PPM at a period of 0.5 days.

\section{Conclusions}
In this paper we have used K2 data, taken during campaigns 3 and 6, to produce and analyse lightcurves for the WASP-75b and WASP-55b systems respectively.
We have refined the orbital parameters of both systems, as the high quality lightcurves allowed us to model the transit with a much greater precision than is possible from ground-based observations. 

Generally, the refined parameters agreed well with the previously published results and we found no major discrepancies between the different sets of data for each planet.
For WASP-55b the parameters produced by this work, shown in Table \ref{tab:w55mcmcparams}, agreed very well with those of \citet{2016MNRAS.457.4205S}.
We found minor differences between our results and those of \citet{2012MNRAS.426..739H} which have arisen as we used the stellar metallicity from \citet{2013AA...558A.106M} rather than CORALIE spectra and we used a different method to estimate the stellar mass.

For the WASP-75b dataset there were minor differences between the values of orbital period, transit depth, transit duration and the impact parameter that we derived and those of \citet{2013AA...559A..36G} that occur as we are able to better constrain the shape of the transit with higher quality data.

We found no evidence of transit timing or duration variations in both systems, and placed upper limits of 25 s and 35 s for TTVs and 100 s and 120 s for TDVs for WASP-55b and WASP-75b respectively. 
There was no evidence of starspot occultations in either system, but we did find low-frequency variation in WASP-75b.
From this we have tentatively estimated its rotation period as $11.24\pm1.54$ days and calculated an estimate of the stellar line-of-sight inclination angle of $i_*=41\pm16^{\circ}$.
We also found no evidence of phase curve variations or additional transiting planets in either system.

\section{ACKNOWLEDGEMENTS}
{\scriptsize
We gratefully acknowledge financial support from the Science and Technology Facilities Council, under grants ST/J001384/1, ST/M001040/1 and ST/M50354X/1. This paper includes data collected by the \textit{K2} mission. Funding for the \textit{K2} mission is provided by the NASA Science Mission directorate. This work made use of {\scriptsize{PyKE}} \citep{pykeref}, a software package for the reduction and analysis of \textit{Kepler} data. This open source software project is developed and distributed by the NASA Kepler Guest Observer Office. This research has made use of the NASA Exoplanet Archive, which is operated by the California Institute of Technology, under contract with the National Aeronautics and Space Administration under the Exoplanet Exploration Program. We also thank the referee for their helpful comments on the previous version of this publication.
}

\bibliographystyle{mnras}
\bibliography{refs.tex}

\begin{thebibliography}{}
\makeatletter
\relax
\def\mn@urlcharsother{\let\do\@makeother \do\$\do\&\do\#\do\^\do\_\do\%\do\~}
\def\mn@doi{\begingroup\mn@urlcharsother \@ifnextchar [ {\mn@doi@}
  {\mn@doi@[]}}
\def\mn@doi@[#1]#2{\def\@tempa{#1}\ifx\@tempa\@empty \href
  {http://dx.doi.org/#2} {doi:#2}\else \href {http://dx.doi.org/#2} {#1}\fi
  \endgroup}
\def\mn@eprint#1#2{\mn@eprint@#1:#2::\@nil}
\def\mn@eprint@arXiv#1{\href {http://arxiv.org/abs/#1} {{\tt arXiv:#1}}}
\def\mn@eprint@dblp#1{\href {http://dblp.uni-trier.de/rec/bibtex/#1.xml}
  {dblp:#1}}
\def\mn@eprint@#1:#2:#3:#4\@nil{\def\@tempa {#1}\def\@tempb {#2}\def\@tempc
  {#3}\ifx \@tempc \@empty \let \@tempc \@tempb \let \@tempb \@tempa \fi \ifx
  \@tempb \@empty \def\@tempb {arXiv}\fi \@ifundefined
  {mn@eprint@\@tempb}{\@tempb:\@tempc}{\expandafter \expandafter \csname
  mn@eprint@\@tempb\endcsname \expandafter{\@tempc}}}

\bibitem[\protect\citeauthoryear{{Agol}, {Steffen}, {Sari}  \&
  {Clarkson}}{{Agol} et~al.}{2005}]{2005MNRAS.359..567A}
{Agol} E.,  {Steffen} J.,  {Sari} R.,   {Clarkson} W.,  2005, \mn@doi [\mnras]
  {10.1111/j.1365-2966.2005.08922.x}, \href
  {http://adsabs.harvard.edu/abs/2005MNRAS.359..567A} {359, 567}

\bibitem[\protect\citeauthoryear{{Aigrain}, {Parviainen}  \& {Pope}}{{Aigrain}
  et~al.}{2016}]{2016MNRAS.459.2408A}
{Aigrain} S.,  {Parviainen} H.,   {Pope} B.~J.~S.,  2016, \mn@doi [\mnras]
  {10.1093/mnras/stw706}, \href
  {http://adsabs.harvard.edu/abs/2016MNRAS.459.2408A} {459, 2408}

\bibitem[\protect\citeauthoryear{{Anderson} et~al.,}{{Anderson}
  et~al.}{2015}]{2015AA...575A..61A}
{Anderson} D.~R.,  et~al., 2015, \mn@doi [\aap] {10.1051/0004-6361/201423591},
  \href {http://adsabs.harvard.edu/abs/2015A%26A...575A..61A} {575, A61}

\bibitem[\protect\citeauthoryear{{Becker}, {Vanderburg}, {Adams}, {Rappaport}
  \& {Schwengeler}}{{Becker} et~al.}{2015}]{2015ApJ...812L..18B}
{Becker} J.~C.,  {Vanderburg} A.,  {Adams} F.~C.,  {Rappaport} S.~A.,
  {Schwengeler} H.~M.,  2015, \mn@doi [\apjl] {10.1088/2041-8205/812/2/L18},
  \href {http://adsabs.harvard.edu/abs/2015ApJ...812L..18B} {812, L18}

\bibitem[\protect\citeauthoryear{{Campante} et~al.,}{{Campante}
  et~al.}{2016}]{2016ApJ...819...85C}
{Campante} T.~L.,  et~al., 2016, \mn@doi [\apj] {10.3847/0004-637X/819/1/85},
  \href {http://adsabs.harvard.edu/abs/2016ApJ...819...85C} {819, 85}

\bibitem[\protect\citeauthoryear{{Christian} et~al.,}{{Christian}
  et~al.}{2006}]{2006MNRAS.372.1117C}
{Christian} D.~J.,  et~al., 2006, \mn@doi [\mnras]
  {10.1111/j.1365-2966.2006.10913.x}, \href
  {http://adsabs.harvard.edu/abs/2006MNRAS.372.1117C} {372, 1117}

\bibitem[\protect\citeauthoryear{{Collier Cameron} et~al.,}{{Collier Cameron}
  et~al.}{2006}]{2006MNRAS.373..799C}
{Collier Cameron} A.,  et~al., 2006, \mn@doi [\mnras]
  {10.1111/j.1365-2966.2006.11074.x}, \href
  {http://adsabs.harvard.edu/abs/2006MNRAS.373..799C} {373, 799}

\bibitem[\protect\citeauthoryear{{Collier Cameron} et~al.,}{{Collier Cameron}
  et~al.}{2007}]{2007MNRAS.380.1230C}
{Collier Cameron} A.,  et~al., 2007, \mn@doi [\mnras]
  {10.1111/j.1365-2966.2007.12195.x}, \href
  {http://adsabs.harvard.edu/abs/2007MNRAS.380.1230C} {380, 1230}

\bibitem[\protect\citeauthoryear{{Daemgen}, {Hormuth}, {Brandner}, {Bergfors},
  {Janson}, {Hippler}  \& {Henning}}{{Daemgen}
  et~al.}{2009}]{2009A&A...498..567D}
{Daemgen} S.,  {Hormuth} F.,  {Brandner} W.,  {Bergfors} C.,  {Janson} M.,
  {Hippler} S.,   {Henning} T.,  2009, \mn@doi [\aap]
  {10.1051/0004-6361/200810988}, \href
  {http://adsabs.harvard.edu/abs/2009A%26A...498..567D} {498, 567}

\bibitem[\protect\citeauthoryear{{Doyle}, {Davies}, {Smalley}, {Chaplin}  \&
  {Elsworth}}{{Doyle} et~al.}{2014}]{2014MNRAS.444.3592D}
{Doyle} A.~P.,  {Davies} G.~R.,  {Smalley} B.,  {Chaplin} W.~J.,   {Elsworth}
  Y.,  2014, \mn@doi [\mnras] {10.1093/mnras/stu1692}, \href
  {http://adsabs.harvard.edu/abs/2014MNRAS.444.3592D} {444, 3592}

\bibitem[\protect\citeauthoryear{{Esteves}, {De Mooij}  \&
  {Jayawardhana}}{{Esteves} et~al.}{2013}]{2013ApJ...772...51E}
{Esteves} L.~J.,  {De Mooij} E.~J.~W.,   {Jayawardhana} R.,  2013, \mn@doi
  [\apj] {10.1088/0004-637X/772/1/51}, \href
  {http://adsabs.harvard.edu/abs/2013ApJ...772...51E} {772, 51}

\bibitem[\protect\citeauthoryear{{Evans} et~al.,}{{Evans}
  et~al.}{2016}]{2016AA...589A..58E}
{Evans} D.~F.,  et~al., 2016, \mn@doi [\aap] {10.1051/0004-6361/201527970},
  \href {http://adsabs.harvard.edu/abs/2016A%26A...589A..58E} {589, A58}

\bibitem[\protect\citeauthoryear{{G{\'o}mez Maqueo Chew} et~al.,}{{G{\'o}mez
  Maqueo Chew} et~al.}{2013}]{2013AA...559A..36G}
{G{\'o}mez Maqueo Chew} Y.,  et~al., 2013, \mn@doi [\aap]
  {10.1051/0004-6361/201322314}, \href
  {http://adsabs.harvard.edu/abs/2013A%26A...559A..36G} {559, A36}

\bibitem[\protect\citeauthoryear{{Groot}}{{Groot}}{2012}]{2012ApJ...745...55G}
{Groot} P.~J.,  2012, \mn@doi [\apj] {10.1088/0004-637X/745/1/55}, \href
  {http://adsabs.harvard.edu/abs/2012ApJ...745...55G} {745, 55}

\bibitem[\protect\citeauthoryear{{Hellier} et~al.,}{{Hellier}
  et~al.}{2009}]{2009Natur.460.1098H}
{Hellier} C.,  et~al., 2009, \mn@doi [\nat] {10.1038/nature08245}, \href
  {http://adsabs.harvard.edu/abs/2009Natur.460.1098H} {460, 1098}

\bibitem[\protect\citeauthoryear{{Hellier} et~al.,}{{Hellier}
  et~al.}{2011}]{2011AA...535L...7H}
{Hellier} C.,  et~al., 2011, \mn@doi [\aap] {10.1051/0004-6361/201117081},
  \href {http://adsabs.harvard.edu/abs/2011A%26A...535L...7H} {535, L7}

\bibitem[\protect\citeauthoryear{{Hellier} et~al.,}{{Hellier}
  et~al.}{2012}]{2012MNRAS.426..739H}
{Hellier} C.,  et~al., 2012, \mn@doi [\mnras]
  {10.1111/j.1365-2966.2012.21780.x}, \href
  {http://adsabs.harvard.edu/abs/2012MNRAS.426..739H} {426, 739}

\bibitem[\protect\citeauthoryear{{Holman} \& {Murray}}{{Holman} \&
  {Murray}}{2005}]{2005Sci...307.1288H}
{Holman} M.~J.,  {Murray} N.~W.,  2005, \mn@doi [Science]
  {10.1126/science.1107822}, \href
  {http://adsabs.harvard.edu/abs/2005Sci...307.1288H} {307, 1288}

\bibitem[\protect\citeauthoryear{{Howell} et~al.,}{{Howell}
  et~al.}{2014}]{2014PASP..126..398H}
{Howell} S.~B.,  et~al., 2014, \mn@doi [\pasp] {10.1086/676406}, \href
  {http://adsabs.harvard.edu/abs/2014PASP..126..398H} {126, 398}

\bibitem[\protect\citeauthoryear{{Jackson}, {Lewis}, {Barnes}, {Drake Deming},
  {Showman}  \& {Fortney}}{{Jackson} et~al.}{2012}]{2012ApJ...751..112J}
{Jackson} B.~K.,  {Lewis} N.~K.,  {Barnes} J.~W.,  {Drake Deming} L.,
  {Showman} A.~P.,   {Fortney} J.~J.,  2012, \mn@doi [\apj]
  {10.1088/0004-637X/751/2/112}, \href
  {http://adsabs.harvard.edu/abs/2012ApJ...751..112J} {751, 112}

\bibitem[\protect\citeauthoryear{{Madhusudhan} \& {Burrows}}{{Madhusudhan} \&
  {Burrows}}{2012}]{2012ApJ...747...25M}
{Madhusudhan} N.,  {Burrows} A.,  2012, \mn@doi [\apj]
  {10.1088/0004-637X/747/1/25}, \href
  {http://adsabs.harvard.edu/abs/2012ApJ...747...25M} {747, 25}

\bibitem[\protect\citeauthoryear{{Maxted}, {Serenelli}  \&
  {Southworth}}{{Maxted} et~al.}{2015}]{2015AA...575A..36M}
{Maxted} P.~F.~L.,  {Serenelli} A.~M.,   {Southworth} J.,  2015, \mn@doi [\aap]
  {10.1051/0004-6361/201425331}, \href
  {http://adsabs.harvard.edu/abs/2015A%26A...575A..36M} {575, A36}

\bibitem[\protect\citeauthoryear{{Mazeh} \& {Faigler}}{{Mazeh} \&
  {Faigler}}{2010}]{2010A&A...521L..59M}
{Mazeh} T.,  {Faigler} S.,  2010, \mn@doi [\aap] {10.1051/0004-6361/201015550},
  \href {http://adsabs.harvard.edu/abs/2010A%26A...521L..59M} {521, L59}

\bibitem[\protect\citeauthoryear{{McLaughlin}}{{McLaughlin}}{1924}]{1924ApJ....60...22M}
{McLaughlin} D.~B.,  1924, \mn@doi [\apj] {10.1086/142826}, \href
  {http://adsabs.harvard.edu/abs/1924ApJ....60...22M} {60}

\bibitem[\protect\citeauthoryear{{McQuillan}, {Aigrain}  \&
  {Mazeh}}{{McQuillan} et~al.}{2013}]{2013MNRAS.432.1203M}
{McQuillan} A.,  {Aigrain} S.,   {Mazeh} T.,  2013, \mn@doi [\mnras]
  {10.1093/mnras/stt536}, \href
  {http://adsabs.harvard.edu/abs/2013MNRAS.432.1203M} {432, 1203}

\bibitem[\protect\citeauthoryear{{Mortier}, {Santos}, {Sousa}, {Fernandes},
  {Adibekyan}, {Delgado Mena}, {Montalto}  \& {Israelian}}{{Mortier}
  et~al.}{2013}]{2013AA...558A.106M}
{Mortier} A.,  {Santos} N.~C.,  {Sousa} S.~G.,  {Fernandes} J.~M.,  {Adibekyan}
  V.~Z.,  {Delgado Mena} E.,  {Montalto} M.,   {Israelian} G.,  2013, \mn@doi
  [\aap] {10.1051/0004-6361/201322240}, \href
  {http://adsabs.harvard.edu/abs/2013A%26A...558A.106M} {558, A106}

\bibitem[\protect\citeauthoryear{{Mo{\v c}nik}, {Clark}, {Anderson}, {Hellier}
  \& {Brown}}{{Mo{\v c}nik} et~al.}{2016}]{2016AJ....151..150M}
{Mo{\v c}nik} T.,  {Clark} B.~J.~M.,  {Anderson} D.~R.,  {Hellier} C.,
  {Brown} D.~J.~A.,  2016, \mn@doi [\aj] {10.3847/0004-6256/151/6/150}, \href
  {http://adsabs.harvard.edu/abs/2016AJ....151..150M} {151, 150}

\bibitem[\protect\citeauthoryear{{Mo{\v c}nik}, {Southworth}  \&
  {Hellier}}{{Mo{\v c}nik} et~al.}{2017}]{2017MNRAS.471..394M}
{Mo{\v c}nik} T.,  {Southworth} J.,   {Hellier} C.,  2017, \mn@doi [\mnras]
  {10.1093/mnras/stx1557}, \href
  {http://adsabs.harvard.edu/abs/2017MNRAS.471..394M} {471, 394}

\bibitem[\protect\citeauthoryear{{Pollacco} et~al.,}{{Pollacco}
  et~al.}{2008}]{2008MNRAS.385.1576P}
{Pollacco} D.,  et~al., 2008, \mn@doi [\mnras]
  {10.1111/j.1365-2966.2008.12939.x}, \href
  {http://adsabs.harvard.edu/abs/2008MNRAS.385.1576P} {385, 1576}

\bibitem[\protect\citeauthoryear{{Rossiter}}{{Rossiter}}{1924}]{1924ApJ....60...15R}
{Rossiter} R.~A.,  1924, \mn@doi [\apj] {10.1086/142825}, \href
  {http://adsabs.harvard.edu/abs/1924ApJ....60...15R} {60}

\bibitem[\protect\citeauthoryear{{Seager} \& {Mall{\'e}n-Ornelas}}{{Seager} \&
  {Mall{\'e}n-Ornelas}}{2003}]{2003ApJ...585.1038S}
{Seager} S.,  {Mall{\'e}n-Ornelas} G.,  2003, \mn@doi [\apj] {10.1086/346105},
  \href {http://adsabs.harvard.edu/abs/2003ApJ...585.1038S} {585, 1038}

\bibitem[\protect\citeauthoryear{{Sing}}{{Sing}}{2010}]{SING2010}
{Sing} D.~K.,  2010, \mn@doi [\aap] {10.1051/0004-6361/200913675}, \href
  {http://adsabs.harvard.edu/abs/2010A%26A...510A..21S} {510, A21}

\bibitem[\protect\citeauthoryear{{Southworth} et~al.,}{{Southworth}
  et~al.}{2016}]{2016MNRAS.457.4205S}
{Southworth} J.,  et~al., 2016, \mn@doi [\mnras] {10.1093/mnras/stw279}, \href
  {http://adsabs.harvard.edu/abs/2016MNRAS.457.4205S} {457, 4205}

\bibitem[\protect\citeauthoryear{{Still} \& {Barclay}}{{Still} \&
  {Barclay}}{2012}]{pykeref}
{Still} M.,  {Barclay} T.,  2012, {PyKE: Reduction and analysis of Kepler
  Simple Aperture Photometry data}, Astrophysics Source Code Library
  (\mn@eprint {ascl} {1208.004})

\bibitem[\protect\citeauthoryear{{Triaud}}{{Triaud}}{2017}]{2017arXiv170906376T}
{Triaud} A.~H.~M.~J.,  2017, preprint, \href
  {http://adsabs.harvard.edu/abs/2017arXiv170906376T} {} (\mn@eprint {arXiv}
  {1709.06376})

\bibitem[\protect\citeauthoryear{{Vanderburg} \& {Johnson}}{{Vanderburg} \&
  {Johnson}}{2014}]{2014PASP..126..948V}
{Vanderburg} A.,  {Johnson} J.~A.,  2014, \mn@doi [\pasp] {10.1086/678764},
  \href {http://adsabs.harvard.edu/abs/2014PASP..126..948V} {126, 948}

\bibitem[\protect\citeauthoryear{{Welsh}, {Orosz}, {Seager}, {Fortney},
  {Jenkins}, {Rowe}, {Koch}  \& {Borucki}}{{Welsh}
  et~al.}{2010}]{2010ApJ...713L.145W}
{Welsh} W.~F.,  {Orosz} J.~A.,  {Seager} S.,  {Fortney} J.~J.,  {Jenkins} J.,
  {Rowe} J.~F.,  {Koch} D.,   {Borucki} W.~J.,  2010, \mn@doi [\apjl]
  {10.1088/2041-8205/713/2/L145}, \href
  {http://adsabs.harvard.edu/abs/2010ApJ...713L.145W} {713, L145}

\makeatother
\end{thebibliography}

\end{document}